\documentclass{rspublic}
\usepackage{psfig}

\newcommand{\R}{{\mathbb R}}

\begin{document}

\title[A new test for chaos]{A new test for chaos
\\ in deterministic systems}

\author[G. Gottwald and I. Melbourne]{Georg A. Gottwald \thanks{School of Mathematics and Statistics, University of Sydney,
NSW 2006, Australia} and Ian Melbourne \thanks{Department of Mathematics and Statistics, University of Surrey,
Guildford, Surrey GU2 7XH, UK}}


\label{firstpage}

\maketitle

\begin{abstract}{Chaos, deterministic dynamical systems, Lyapunov exponents, 
mean square displacement, Euclidean extension}
We describe a new test for determining whether a given
deterministic dynamical system is chaotic or nonchaotic.
In contrast to the usual method of computing the maximal Lyapunov exponent,
our method is applied directly to the time series data and does not require
phase space reconstruction.
Moreover, the dimension of the dynamical system and the form of
the underlying equations is irrelevant.  
The input is the time series data and the output is $0$ or $1$ depending on
whether the dynamics is non-chaotic or chaotic.
The test is universally applicable to any deterministic dynamical system,
in particular to ordinary and partial differential equations, and to maps.

Our diagnostic is the real valued function
$p(t)=\int_0^t \phi({\bf x}(s))\cos(\theta(s))\rd s$ where $\phi$
is an observable on the underlying dynamics ${\bf x}(t)$ and 
$\theta(t)=ct+\int_0^t\phi({\bf x}(s))\rd s$.
The constant $c>0$ is fixed arbitrarily.
We define the mean-square-displacement ${\bf M}(t)$ for $p(t)$ and set
$K=\lim_{t \rightarrow \infty} \log{\bf M}(t)/\log{t}$.
Using recent developments in ergodic theory, we argue that typically $K=0$
signifying nonchaotic dynamics or $K=1$
signifying chaotic dynamics.
\end{abstract}

\section{\label{sec:intro}Introduction}

The usual test of whether a deterministic dynamical system is chaotic or 
nonchaotic is the calculation of the largest Lyapunov exponent $\lambda$. 
A positive largest Lyapunov exponent indicates chaos:
if $\lambda>0$, then nearby trajectories
separate exponentially and if $\lambda<0$, then nearby
trajectories stay close to each other. 
This approach has been widely used for dynamical systems whose
equations are known (Abarbanel {\em et al.} 1993; Eckmann {\em et al.} 1986;
Parker \& Chua 1989).
If the equations are not known
or one wishes to examine experimental data, this approach is
not directly applicable. However Lyapunov exponents may be estimated
(Wolf {\em et al.} 1985; Sana \& Sawada 1985; Eckmann {\em et al.} 1986; 
Abarbanel {\em et al.} 1993) by using the embedding theory of Takens (1981) or
by approximating the linearisation of the evolution operator.
Nevertheless, the computation of Lyapunov exponents is greatly facilitated
if the underlying equations are known and are low-dimensional.

In this article, we propose a new $0$--$1$ test for chaos which
does not rely on knowing the underlying equations,
and for which the dimension of the equations is irrelevant.  
The input is the time series data and the output is 
either a $0$ or a~$1$ depending on whether the dynamics is nonchaotic
or chaotic.  Since our method
is applied directly to the time series data, the only difference in
difficulty between analysing a system of partial differential
equations or a low-dimensional system of ordinary differential equations
is the effort required to generate sufficient data.  (As with all approaches,
our method is impracticable if there are extremely long transients or once 
the dimension of the attractor becomes too large.)  
With experimental data, there is the additional effect of noise to be 
taken into consideration.   We briefly discuss this important issue at at 
the end of this paper.  However, our aim in this paper is to present our 
findings in the situation of noise-free deterministic data.


\section{\label{sec:test}Description of the $0$--$1$ test for chaos}

To describe the new test for chaos,
we concentrate on the continuous time case and 
denote a solution of the underlying system by ${\bf x}(t)$.  
The discrete time case is handled analogously with the
obvious modifications.
Consider an observable $\phi({\bf{x}})$ of the underlying dynamics. The
method is essentially independent of the actual form of $\phi$ --- almost
any choice of $\phi$ will suffice. For example if
${\bf{x}}=(x_1,x_2,\ldots,x_n)$ then $\phi({\bf{x}})=x_1$ is a
possible and simple choice. 
Choose $c>0$ arbitrarily and define 
\begin{align} \label{p} \nonumber
\theta(t) &= ct+\int_0^t \phi({\bf x}(s))\rd s, \\
p(t) &= \int_0^t \phi({\bf x}(s))\cos(\theta(s))\rd s.
\end{align}
(Throughout the examples in \S\ref{sec:example} and \S\ref{sec:morex}
we fix $c=1.7$ once and for all.)
We claim that 
\begin{itemize}
\item[(i)] $p(t)$ is bounded if the underlying dynamics is
nonchaotic and 
\item[(ii)]
$p(t)$ behaves asymptotically like a Brownian motion
if the underlying dynamics is chaotic.
\end{itemize}

The definition of $p$ in~\eqref{p}, which involves only the observable
$\phi({\bf x})$, highlights the universality of the test.
The origin and nature of the data which is fed into the system~\eqref{p}
is irrelevant for the test, and so is the dimension of the underlying dynamics.

Later on, we briefly explain the justification behind the claims (i) and (ii).
For the moment, we suppose that the claims are correct and show how to proceed.

To characterise the growth of the function $p(t)$ defined in~\eqref{p},
it is natural to look at
the mean square displacement (MSD) of $p(t)$, defined to be
\begin{align}\label{MSD}
{\bf M}(t)
=
\lim_{T\rightarrow \infty}
\frac1T\int_0^T\left( p(t+\tau)-p(\tau) \right)^2 \rd\tau.
\end{align}
If the behaviour of $p(t)$ is Brownian, i.e.\ the underlying dynamics is
chaotic, then $M(t)$ grows linearly in time; if the
behaviour is bounded, i.e.\ the underlying dynamics is nonchaotic,
then $M(t)$ is bounded.  We define 
$K=\lim_{t \rightarrow \infty} \log{\bf M}(t)/\log{t}$ as the asymptotic
growth rate of the MSD.  The growth rate $K$ can be
numerically determined by means of linear regression of $\log{\bf
M}(t)$ versus $\log t$ (Press {\em et al.} 1992). 
(To avoid negative logarithms we
actually calculate $\lim_{t \rightarrow \infty} \log({\bf
M}(t)+1)/\log{t}$ which obviously does not change the slope $K$.) 
This allows for a clear distinction of a nonchaotic and a
chaotic system as $K$ may only take values $K=0$ or $K=1$. We have
lost though the possibility of quantifying the chaos by the magnitude
of the largest Lyapunov exponent $\lambda$.

Numerically one has to make sure that
initial transients have died out so that the
trajectories are on (or close to) the attractor at time zero, and
that the integration time $T$ is long enough to ensure
$T\gg t$.

\section{\label{sec:example}An example: the forced van der Pol oscillator}
We illustrate the $0$--$1$ test for chaos with the help of a concrete example, 
the forced van der Pol system,
\begin{align}\label{vanderpol}
\dot{x_1} &= x_2 \nonumber\\
\dot{x_2} &= -d(x_1^2-1)x_2-x_1+a\cos\omega t
\end{align}
which has been widely studied in nonlinear dynamics
(van der Pol 1927; Guckenheimer \& Holmes 1990).  
For fixed $a$ and $d$, the dynamics may be chaotic or nonchaotic depending on 
the parameter $\omega$.
Following Parlitz \& Lauterborn (1987), we take $a=d=5$ and let $\omega$ 
vary from $2.457$ to $2.466$ in increments of $0.00001$.
Choose $\phi ({ \bf x})=x_1+x_2$ and $c=1.7$. 
We stress that the results are independent of these choices for all 
practical purposes.
As described below in \S\ref{sec:just}, almost all choices will 
work.
(Deliberately poor choices such as $c=0$, or $\phi=7$ or $\phi=t$ obviously 
fail; sensible choices that fail are virtually impossible to find.)

\begin{figure}[htb]
\centerline{%
\psfig{file=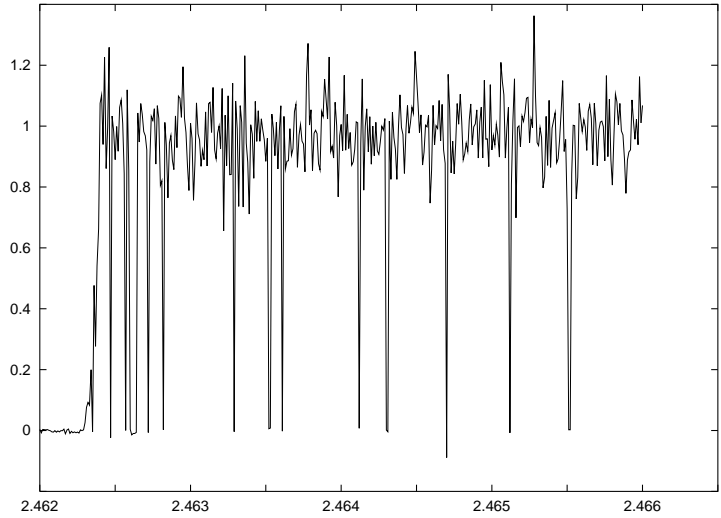,angle=270,width=4.0in}}
\caption{\label{fig:skew} Asymptotic growth rate $K$ of the mean square 
displacement~\eqref{MSD} versus
$\omega$ for the van der Pol system~\eqref{vanderpol} determined by a
numerical simulation of the skew product system~\eqref{vanderpol} and~\eqref{p} 
with $a=d=5$, $c=1.7$, $\phi({\bf x})=x_1+x_2$
 and $\omega$ varying from 2.462 to 2.466.  
The integration interval is $T=2,000,000$ after an initial transient 
of $200,000$ units of time.}
\end{figure}

\begin{figure}[htb]
\centerline{%
\psfig{file=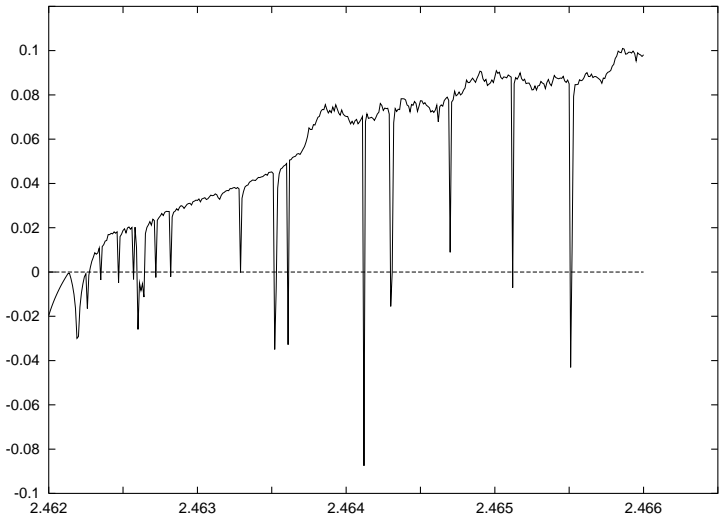,angle=270,width=4.0in}}
\caption{\label{fig:Lyap} Largest Lyapunov exponent $\lambda$ versus
$\omega$ for the van der Pol system~\eqref{vanderpol} with
$a=d=5$ and $\omega$ varying from 2.462 to 2.466 
(cf Parlitz \& Lauterborn 1987).
The integration interval is $T=2,000,000$ after an initial transient of 
$200,000$ units of time.}
\end{figure}

In figure~\ref{fig:skew} we show a plot of $K$ versus $\omega$. The
periodic windows are clearly seen. 
As a comparison we show in 
figure~\ref{fig:Lyap} the largest Lyapunov exponent $\lambda$ versus $\omega$. 
Since the onset of chaos does not occur until after $\omega=2.462$ 
we display the results only for the range $2.462<\omega<2.466$ in 
figures~\ref{fig:skew} and~\ref{fig:Lyap}.
(Both methods easily indicate regular dynamics for $2.457<\omega<2.462$.)

\begin{figure}[htb]
\centerline{%
\psfig{file=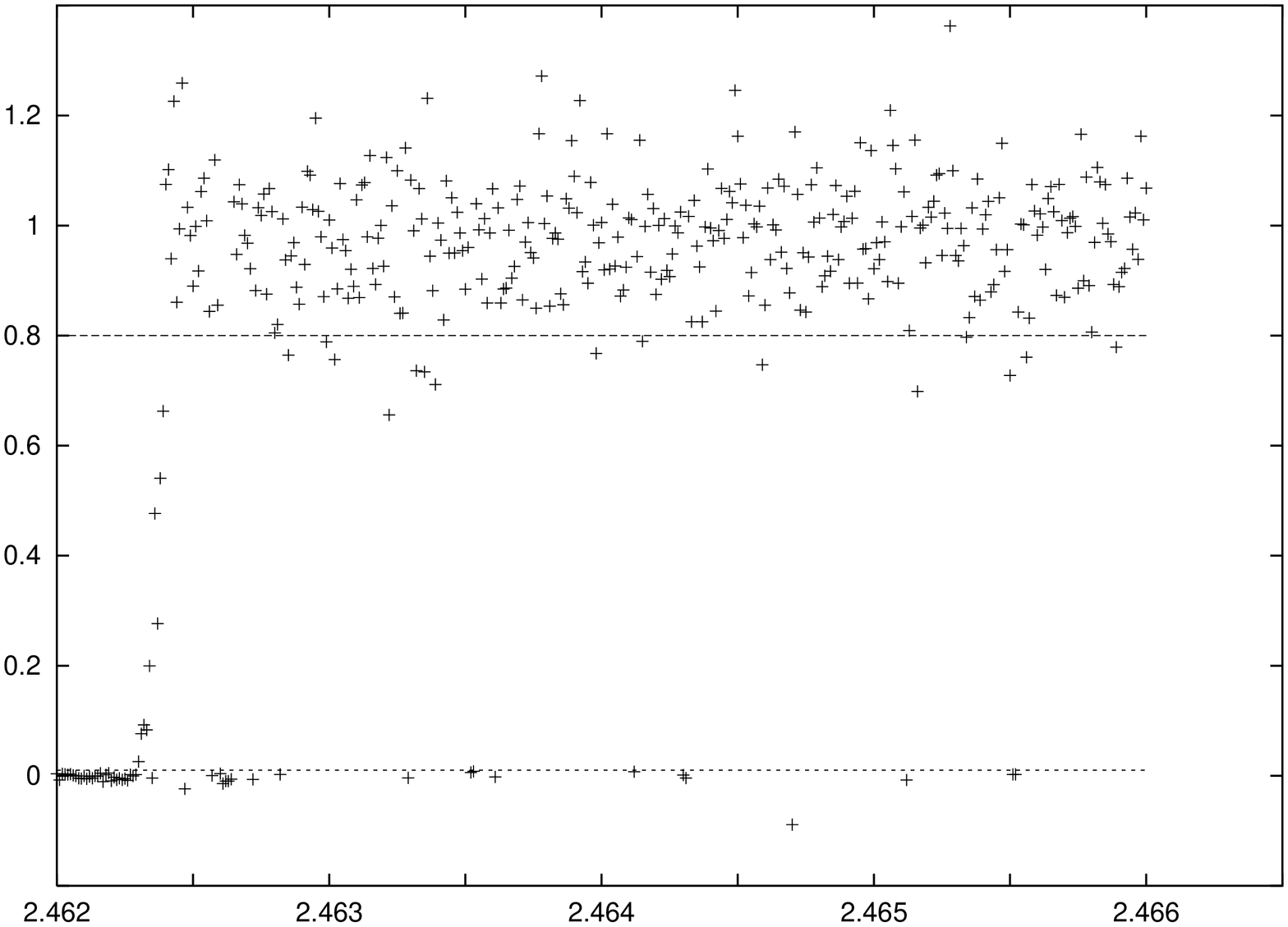,angle=270,width=4.0in}}
\caption{\label{fig:cutoff} Asymptotic growth rate $K$ 
versus $\omega$ for the van der Pol system~\eqref{vanderpol} 
as in figure~\ref{fig:skew} with $T=2,000,000$.  
The horizontal lines represent $K=0.01$ and $K=0.8\,$.}
\end{figure}

Naturally we do not obtain the values $K=0$ and $K=1$ exactly -- the 
mathematical results that underpin our method predict these values 
in the limit of infinite integration time.   
(The same caveat applies equally to the Lyapunov exponent method.)
In producing the data
for figures~\ref{fig:skew} and~\ref{fig:Lyap}, we allowed for a transient
of 200,000 units of time and then integrated up to time $T=2,000,000$.
As can be seen in figure~\ref{fig:cutoff}, for most of the $400$ data points
in the range of $\omega$, we obtain $K>0.8$ or $K<0.01$.


\begin{figure}[thb]
\centerline{%
\psfig{file=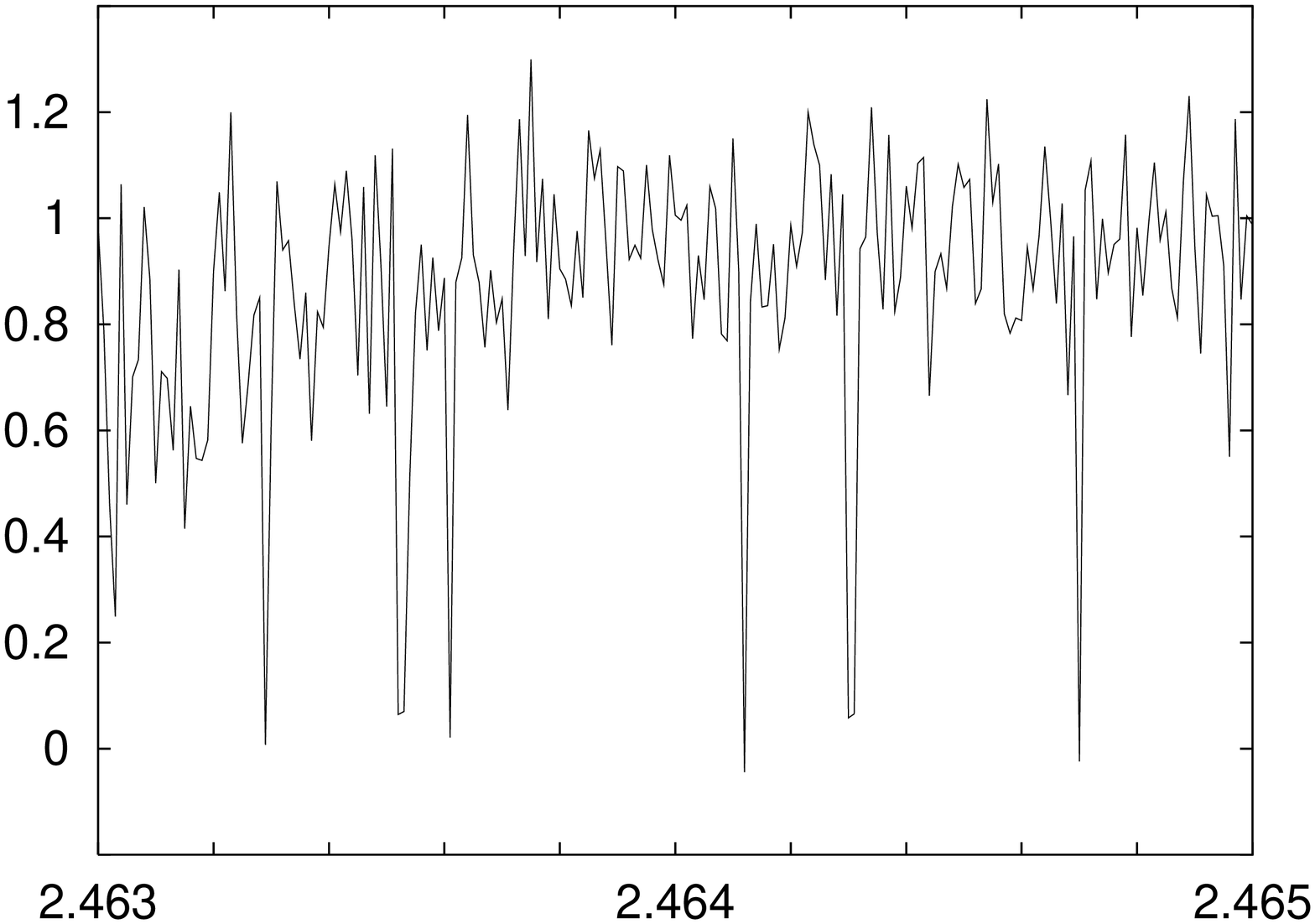,angle=270,width=2.5in}
\psfig{file=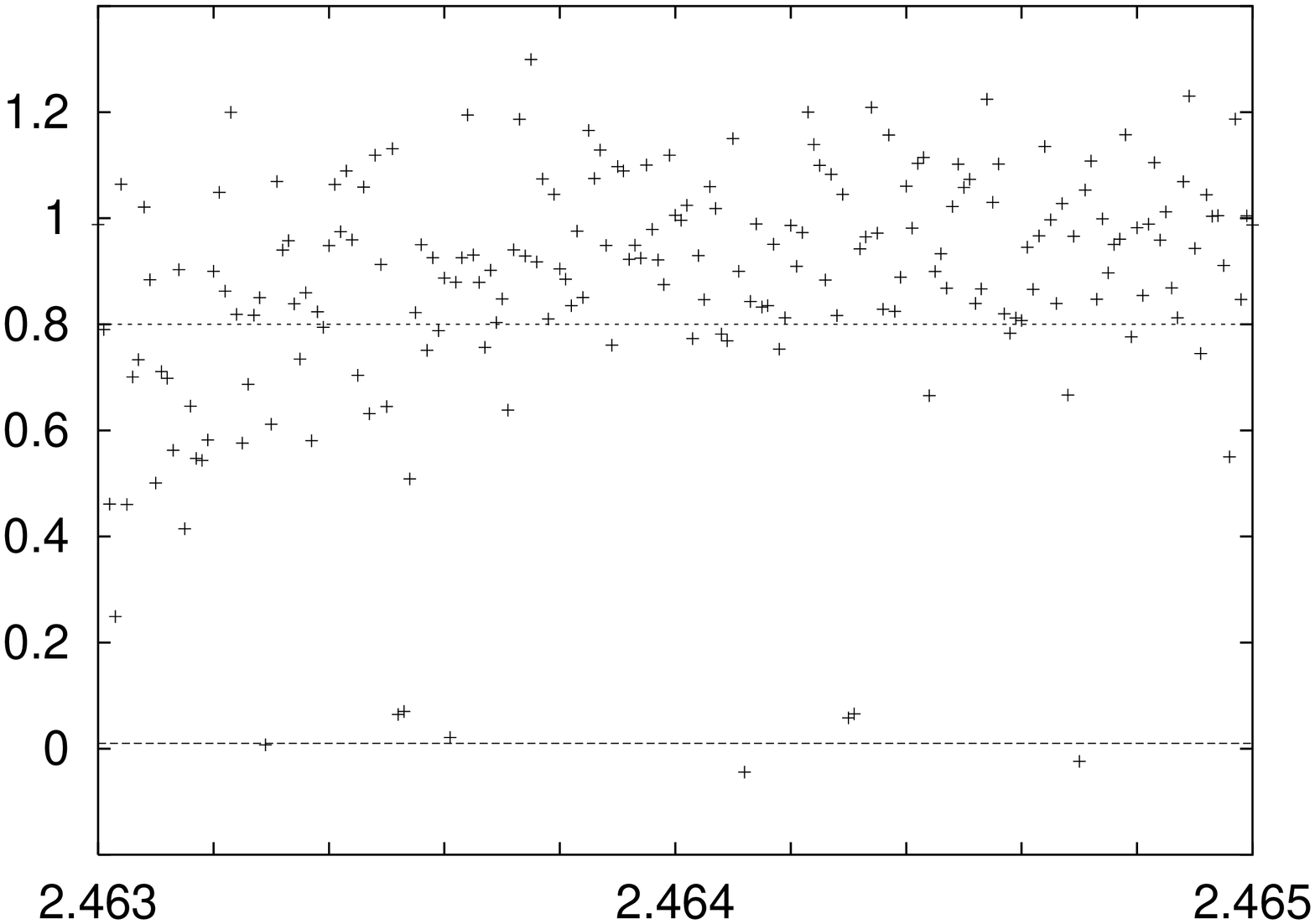,angle=270,width=2.5in}}
\caption{\label{fig:skew4} Asymptotic growth rate $K$ 
versus $\omega$ varying from 2.463 to 2.465
for the van der Pol system
as in figures~\ref{fig:skew} and~\ref{fig:cutoff} but with
integration interval $T=50,000$ after an initial transient of $200,000$ 
units of time.
The horizontal lines represent $K=0.01$ and $K=0.8\,$.}
\end{figure}

Next, we carry out the $0$--$1$ test for the forced
van der Pol system in the situation of a more limited quantity of data.
The results are shown in figure~\ref{fig:skew4} for $2.463<\omega<2.465$.
We again allow for a transient time 200,000 but then integrate only
for $T=50,000$.   
The transitions between chaotic dynamics and periodic windows are almost as
clear with $T=50,000$ as they are with $T=2,000,000$ even though the convergence
of $K$ to $0$ or $1$ is better with $T=2,000,000$.

\clearpage

\section{\label{sec:morex}Further examples}

 To test the method on high-dimensional systems we investigated the
driven and damped Kortweg-de~Vries (KdV) equation (Kawahara \& Toh 1988)
\begin{align} \label{Toh}
u_t+uu_x+\beta u_{xxx}+\alpha u_{xx}+\nu u_{xxxx}=0,
\end{align}
on the interval $[0,40]$ with periodic boundary conditions.
This partial differential equation has non-chaotic solutions if the
dispersion $\beta$ is 
large and exhibits spatio-temporal chaos for sufficiently small $\beta$.
Note that equation~\eqref{Toh} reduces to the KdV equation when $\alpha=\nu=0$,
and reduces to the Kuramoto-Sivashinksy equation when $\beta=0$.

We fix $\alpha=2$, $\nu=0.1$ and vary $\beta$.
For $\beta=0$, it is expected that the dynamics of the Kuramoto-Sivashinksy 
equation are chaotic for these parameter values.
As an observable we used $\phi(u(x,t))=u(x_0,t)$ where $x_0$  is an arbitrarily fixed position, and we iterated until time $T=35,000$.
The $0$--$1$ test confirms that the dynamics is chaotic at $\beta=0$ 
(with $K=0.939$).   Also, we obtain $K=0.989$ at $\beta=0.1$ and $K=0.034$ 
at $\beta=4$, indicating chaotic and regular dynamics respectively at these two
parameter values.

Finally, for discrete dynamical systems, we tried out the test with an 
ecological model whose 
chaotic component is coupled with strong periodicity 
(Brahona \& Poon 1996; Cazelles \& Ferriere 1992).  The model
\begin{align*}
x_{k+1} & = 118\, y_k\exp(-0.001(x_k+y_k)) \\
y_{k+1} & = 0.2\, x_k\exp(-0.07(x_k+y_k))
+ 0.8\, y_k\exp(-0.05(0.5\, x_k+y_k))
\end{align*}
has a non-connected chaotic attractor consisting of seven connected components.
Our test yields $K=1.023$ with only $10,000$ data points and clearly shows that
the dynamics is chaotic.

\section{\label{sec:just}Justification of the $0$--$1$ test for chaos}

The function $p(t)$ can be viewed as a component of the solution to the skew
product system
\begin{align}\label{skew}
\dot{\theta}&=c+\phi({\bf{x}}(t)) \nonumber \\
\dot{p}&=\phi({\bf{x}}(t))\cos{\theta} \\
\nonumber
\dot{q}&=\phi({\bf{x}}(t))\sin{\theta}
\end{align}
driven by the dynamics of the observable $\phi({\bf x}(t))$.
Here $(\theta,p,q)$ represent coordinates on the Euclidean group
${\bf E}(2)$ of rotations $\theta$ and translations $(p,q)$
in the plane.

\begin{figure}
\centerline{
\psfig{file=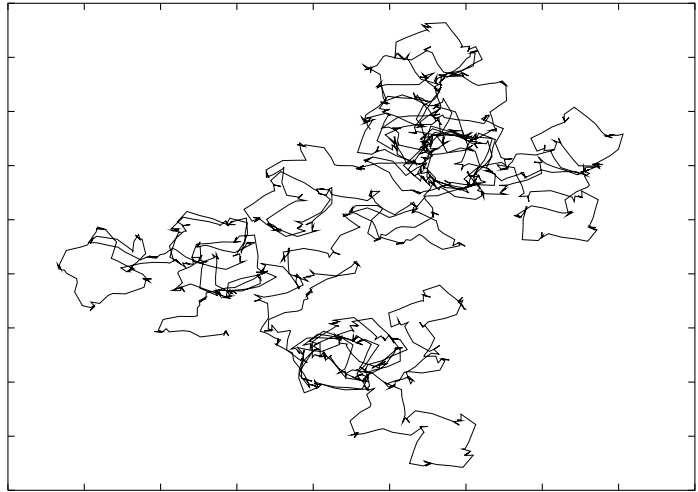,angle=270,width=4.0in}}

\centerline{(a)}

 \vspace{.3in}
\centerline{%
\psfig{file=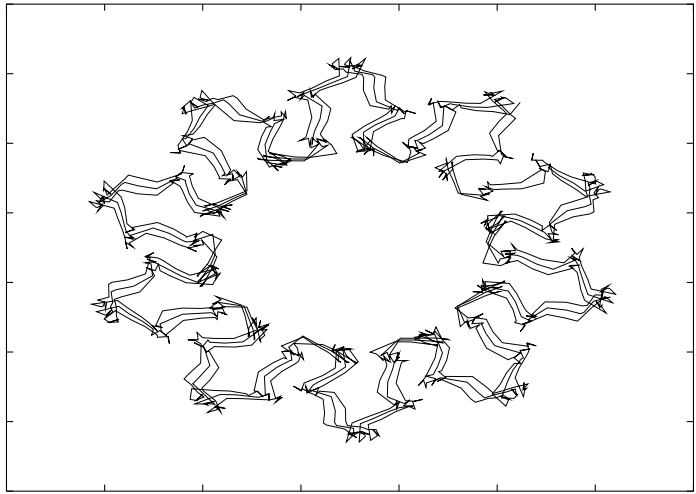,angle=270,width=4.0in}} 

\centerline{(b)}
 \vspace{.3in}


\caption{\label{fig:dynam}
The dynamics of the translation components $(p,q)$ of the
${\bf E}(2)$-extension~\eqref{skew} for the van der Pol system~\eqref{vanderpol}
with $a=d=5$, $c=1.7$, $\phi({\bf x})=x_1+x_2$.
These plots were obtained by integrating for $T\sim 1400$ 
(with timestep $0.01$).
In~(a), an unbounded trajectory is shown corresponding
to chaotic dynamics at $\omega=2.46550$.
In~(b), a bounded trajectory is shown
corresponding to regular dynamics at $\omega=2.46551$.
}

\end{figure}

We note that inspection of the dynamics of the $(p,q)$-trajectories of the group
extension provides very quickly (for small $T$) a simple visual test of 
whether the underlying dynamics is chaotic or nonchaotic as can be seen 
from figure~\ref{fig:dynam}. 
\clearpage

\clearpage

In Nicol {\em et al.} (2001) it has been shown that 
typically the dynamics on the group extension 
is sublinear and furthermore that
typically the dynamics is bounded if the underlying dynamics is nonchaotic and 
unbounded (but sublinear) if the underlying dynamics is chaotic.
Moreover, the $p$ and $q$ components
each behave like a Brownian motion on the line if the chaotic attractor is
uniformly hyperbolic (Field {\em et al.} 2002).  A nondegeneracy result of
Nicol {\em et al.} (2001) ensures
that the variance of the Brownian motion is nonzero for almost all
choices of observable $\phi$.  Recent work of Melbourne \& Nicol (2002) 
indicates that these statements
remain valid for large classes of nonuniformly hyperbolic systems, such as
H\'enon-like attractors.    

There is a weaker condition on the `chaoticity' of $X$ that guarantees the 
desired growth rate 
$K=1$ for the mean square displacement~\eqref{MSD}: namely that the 
autocorrelation function of $\phi({\bf x})\cos(\theta)$ decays at a
rate that is better than quadratic.  More precisely, 
let ${\bf x}(t)$ and $\theta(t)$ denote solutions to the skew product
equations~\eqref{skew}
with initial conditions ${\bf x}_0$ and $\theta_0$ respectively.
If there are constants $C>0$ and $\alpha>2$ such that
\[
\left|\int \phi({\bf x}(t))\cos(\theta(t))\phi({\bf x}_0)\cos\theta_0\rd{\bf x}_0 
\rd\theta_0 \right|\le Ct^{-\alpha},
\]
for all $t>0$, then $K=1$ as desired 
(Biktashev \& Holden 1998; Ashwin {\em et al.} 2001; Field {\em et al.} 2002).
(Again, results of Nicol {\em et al.} (2001); Field {\em et al.} (2002); and
Melbourne \& Nicol (2002) ensure that the appropriate nondegeneracy
condition holds for almost all choices of $\phi$.)
There is a vast literature on proving decay of correlations (Baladi 1999) 
and this has been generalised to the equivariant setting for discrete time by
Field {\em et al.} (2002) and Melbourne \& Nicol (2002)
and for continuous time by Melbourne \& T\"or\"ok (2002).  
It follows from these references that $K=1$, for large classes of 
chaotic dynamical systems.

One might ask why it is not better to work, instead of the ${\bf E}(2)$-extension,
with the simpler $\R$-extension 
\[
\dot p=\phi({\bf x}(t).
\]
In principle, $p(t)$ can again be used as a detector for chaos.
However, by the ergodic theorem $p(t)$ will typically grow linearly
with rate equal to the space average of $\phi$.
This would lead to $K=2$ irrespective of whether the dynamics is regular or 
chaotic.  Hence, it is necessary to subtract off the linear term of $p(t)$ in 
order to observe the bounded/diffusive growth that distinguishes between 
regular/chaotic dynamics.   Subtracting this linear term is a highly nontrivial
numerical obstruction.  The inclusion of the $\theta$ variable in the 
definition~\eqref{p} of $p(t)$ and in the skew product~\eqref{skew}
kills off the linear term.  


\section{\label{sec:discuss_potato}Discussion}
We have established a simple, inexpensive, and novel $0$--$1$ test for chaos.
The computational effort is
of low cost, both in terms of programming efforts and in terms of
actual computation time. The test is a binary test in the sense that
it can only distinguish between nonchaotic and chaotic dynamics. This
distinction is extremely clear by means of the diagnostic variable $K$
which has values either close to $0$ or close to $1$. The most
powerful aspect of our method is that it is independent of the nature
of the vector field (or data) under consideration.  In particular the
equations of the underlying dynamical system do not need to be known,
and there is no practical restriction on the dimension of the
underlying vector field. 
In addition, our method applies to the observable $\phi({\bf x}(t))$
rather than the full trajectory ${\bf x}(t)$.

Related ideas (though not with the aim
to detect chaos) have been used for PDE's in the context of
demonstrating hypermeander of spirals in excitable media 
(Biktashev \& Holden 1998)
where the spiral tip appears to undergo a planar Brownian motion
(see also Ashwin {\em et al.} 2001). 
We note also the work of Coullet \& Emilsson (1996) who studied the dynamics 
of Ising walls
on a line, where the motion is the superposition of a linear drift and 
Brownian motion.  (This is an example of an $\R$-extension which we mentioned
briefly in \S\ref{sec:just}.  As we pointed out then, the linear drift
is an obstruction to using an $\R$-extension to detect chaos.)

From a purely computational point of view, 
the method presented here has a number of advantages over the
conventional methods using Lyapunov exponents.
At a more technical level, we note that the computation of Lyapunov 
exponents can be thought of abstractly
as the study of the 
${\bf GL}(n)$-extension 
\begin{align*}
\dot {\bf A} = (d{\bf f})_{{\bf x}(t)}{\bf A}
\end{align*}
where ${\bf GL}(n)$ is the space of $n\times n$ matrices, 
${\bf A}\in{\bf GL}(n)$, and $n$ is the size of the
system.   Note that the extension involves $n^2$ additional equations and
is defined using the linearisation of the 
dynamical system.  To compute the dominant exponent, it is still necessary to
add $n$ additional equations, again governed by the linearised system.
In contrast, our method requires the addition of two equations.

In this paper, we have concentrated primarily on the idealised situation
where there is an in principle unlimited amount of noise-free data.
However, in \S\ref{sec:example} we also demonstrated the effectiveness of
our method for limited data sets.
An issue which will
be pursued in further work is the effect of noise which is inevitably
present in all experimental time series.
Preliminary results show that our test can cope with small amounts of noise
without difficulty.  A careful study of this capability, and comparison
with other methods, is presently in progress.

\begin{acknowledgements}
We are grateful to Philip Aston, Charlie Macaskill and Trevor Sweeting 
for helpful discussions and suggestions.
The research of GG was supported in part by the European Commission 
funding for the Research Training Network ``Mechanics and Symmetry in Europe''
(MASIE).
\end{acknowledgements}

\label{lastpage}

\end{document}